\documentclass[conference]{IEEEtran}
\IEEEoverridecommandlockouts
\usepackage{cite}
\usepackage{amsmath,amssymb,amsfonts}
\usepackage{algorithmic}
\usepackage{graphicx}
\usepackage{textcomp}
\usepackage{xcolor}
\def\BibTeX{{\rm B\kern-.05em{\sc i\kern-.025em b}\kern-.08em
    T\kern-.1667em\lower.7ex\hbox{E}\kern-.125emX}}
\begin{document}

\title{The effect of inhomogeneous carbon prices on the cost-optimal design of a simplified European power system}

\author{
\IEEEauthorblockN{Markus Schlott, Fabian Hofmann \\ Rosana de Oliveira Gomes, Alexander Kies}
\IEEEauthorblockA{Frankfurt Institute for Advanced Studies \\ Goethe University Frankfurt \\ Frankfurt am Main, Germany}
\and
\IEEEauthorblockN{Changlong Wang}
\IEEEauthorblockA{Australian-German Energy Transition Hub \\ The University of Melbourne \\ Melbourne \\ Australia}
}


\maketitle

\begin{abstract}
Carbon prices are one of the most prominent methods to reduce global greenhouse gas emissions and have been adopted by several countries around the world. However, regionally different carbon prices can lead to carbon leakage. \\
In this paper, a simplified European power system where carbon prices are varied with respect to GDP per capita is investigated. The findings show inhomogeneous carbon prices leading to significant carbon leakage due to coal-fired generation remaining a major source of power in Eastern Europe.
From the results it can be concluded that inhomogeneous carbon prices are a potential risk for European long term decarbonisation targets.
\end{abstract}

\begin{IEEEkeywords}
Carbon Leakage, Carbon Price, European Energy System
\end{IEEEkeywords}

\section{Introduction}
Global warming is the human-caused long-term increase in observed average temperature on Earth.
Power systems around the world are transforming towards high shares of renewable energy sources to mitigate global warming.
Carbon prices are a prominent option to include the externality of greenhouse cases in economic processes. However, varying carbon prices around the globe could lead to carbon leakage. In addition, Weitzman \cite{weitzman2017voting} has shown and Nordhaus \cite{nordhaus2019climate} has pointed out that it is easier to negotiate
a single carbon price than to set different limits on carbon emissions per country.
The concept of carbon leakage is described as the increase of carbon dioxide emissions in a country as a result of lower carbon dioxide emissions in another country, for instance by moving energy-intensive industries from one region to another.
 
Zhu et al. \cite{zhu2019impact} have recently taken a look at carbon taxation in a coupled electricity and heat system for Europe and concluded that a CO$_2$ tax is mandatory for decarbonisation.
However, Patt and Lilliestam \cite{patt2018case} have argued, based on ideas from transitions theory, that carbon pricing is outdated and that different policies are required to curb greenhouse gas emissions.

Multiple estimates on appropriate carbon prices exist. Ackerman and Stanton \cite{ackerman2012climate} have investigated the "social cost of carbon" using the DICE model and have found that ambitious scenarios that reach zero or net negative emissions by the end of the century require prices of 150 to 500 \$ per ton of carbon dioxide reduction. 
In Sweden, a carbon tax has already been a major instrument of climate policy since 1991 and has been considered highly effective \cite{carbontax,johansson2000carbon}. The level of the Swedish carbon tax rose from around 25 Euro per ton in 1991 to around 120 Euro per ton in 2017, although the tax level for the industry has been significantly below the standard level during the entire period.
Besides Sweden, several other European countries have experimented with carbon taxes in the past, but struggled to introduce it uniformly across different economic sectors \cite{andersen2010europe}.

In this paper we investigate a simplified European power system with regionally diversified carbon prices. Carbon prices are assumed to be higher in wealthier regions as reflected by gross domestic product (GDP) per capita than in regions with low GDP per capita.
We show that diversified CO$_2$ prices lead to enormous leakage of carbon emissions to countries with low GDP per capita and this effect is at times even stronger than the general reduction of emissions due to carbon prices.

\section{Methodology}
\begin{figure}[!h]
    \centering
    \includegraphics[width=.5\textwidth]{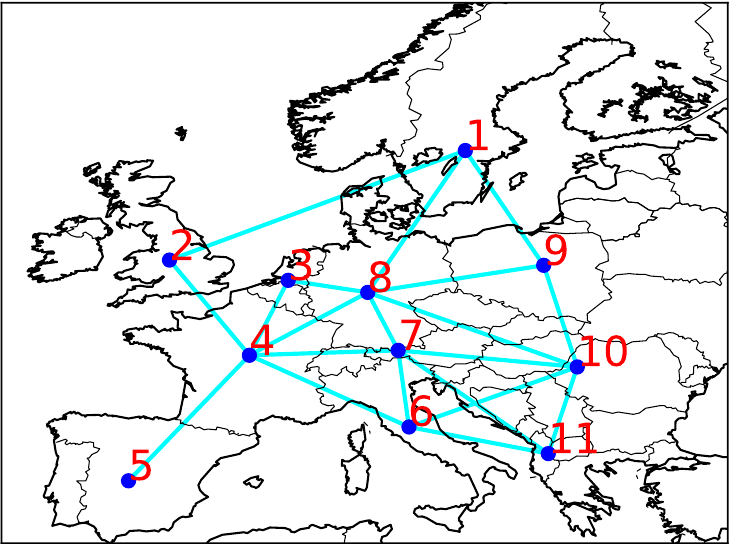}
    \caption{Topology of the investigated system. Nodes are described in Table \ref{tab:nodes}.}
    \label{fig:topology}
\end{figure}
In this paper, a simplified European power system where neighbouring countries with similar GDP per capita are joint into single nodes, is considered. The topology of the system is shown in Fig. \ref{fig:topology}.
Information on the single nodes with relevant quantities are shown in Table \ref{tab:nodes}. They are connected via inter-connecting transmission links.
\begin{table}[!h]
    \centering
    \begin{tabular}{c|c|c|c}
       Node  & Countries & GDP per capita & $\left<d_{n,t}\right>$ \\\hline
        1 & SE/NO/DK/FI & 60.149 & 45.4 GW \\
        2 & IE/GB & 44.869 & 42.1 GW \\
        3 & NL/BE/LU &51.745 & 23.8 GW\\
        4 & FR & 41.463 & 54.3 GW\\
        5 & ES/PT & 29.193 & 34.8 GW\\
        6 & IT & 34.318 & 36.8 GW\\
        7 & AT/CH & 66.877 & 15.0 GW\\
        8 & DE & 48.195 & 59.1 GW\\
        9 & PL/LT/LV/EE & 15.998 & 21.6 GW\\
        10 & RO/BG/HU/CZ/SK & 15.494 & 26.2 GW\\
        11 & GR/SI/HR/RS/AL/BA/ME/XK/MK & 12.842 & 17.5 GW
    \end{tabular}
    \caption{Nodes considered in the model. GDP per capita is given in monetary units [mu].}
    \label{tab:nodes}
\end{table}
We assume carbon prices are not the same for every region, but instead differentiated. 
The carbon price per region $\mu_{\text{CO}_2}$ is defined as a function of GDP per capita and the base carbon price $\bar{\mu}$ via:
\begin{align}
    \mu_{CO_2}\left(\text{GDP}\right) &= \alpha \bar{\mu} \frac{\text{GDP}}{\bar{\text{GDP}}} + \bar{\mu} - \bar{\mu}\alpha \\
    \mu_{CO_2}\left(\text{GDP}\right) &\geq 0
\end{align}
Here, $\bar{\text{GDP}} = \frac{\sum_n \text{GDP}_n \left<d_n\right>}{\sum_n \left<d_n\right>}$ is the demand-weighted average GDP per capita.
The parameter $\alpha$ describes the distribution of carbon prices and is varied in the following together with the base price $\bar{\mu}$.
For an $\alpha < \frac{1}{1-\frac{\text{GDP}^\text{min}_n}{\bar{\text{GDP}}}}$ the second condition is trivially true. In addition, if $\alpha$ is chosen above this maximum value, the demand-weighted carbon price does not equal the carbon base price anymore.

\begin{figure}[!h]
    \centering
    \includegraphics[width=.49\textwidth]{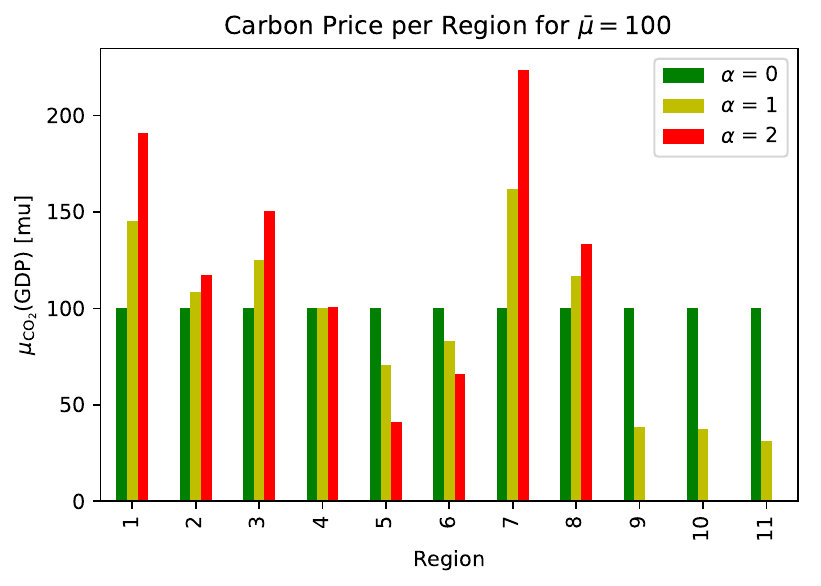}
    \caption{Carbon price per region for a base carbon price of $\bar{\mu}=100$ in dependency of the distribution parameter $\alpha=0$ (green), $\alpha=1$ (yellow) and $\alpha=2$ (red).}
    \label{fig:regionalprice}
\end{figure}

Carbon prices per region for a base price of $100$ mu/ton are shown in Fig. \ref{fig:regionalprice}.
For a strongly inhomogeneous distribution ($\alpha=2$), carbon prices vanish in regions 9-11, whereas they strongly increase to values at or above 150 mu/ton in regions 1,3 and 7.
\subsection{Model}
The model employed to study the problem of carbon leakage in the context of a European power system is a linear minimisation model of total system cost.
The optimisation objective is to find the cost-optimal solution for the simplified power system under various constraints.
At every node, the model can build and use various generators (wind, solar PV, coal, open cycle gas turbine (OCGT)) as well as storage facilities (battery, hydro power) to satisfy the given hourly demand over a single year.
The problem objective reads:
\begin{align}
 \min_{g,G,F} & \left(\sum_{n,s} c_{n,s} \cdot G_{n,s} + \sum_{n,s,t} o_{s} \cdot g_{n,s,t} \right). \label{eq:minimisation}
\end{align}
The objective consists of capital costs $c_{n,s}$ for installed capacity $G_{n,s}$ of an energy carrier $s$ at node $n$ and marginal costs of generation $o_{s}$ for energy generation $g_{n,s,t}$ of carrier $s$ at node $n$ and time $t$.

This optimisation objective is subject to various constraints; most important are the nodal power balance among generation, demand and flow, as well as the dispatch constraints with respect to generation:
\begin{align}
\begin{split}
\sum_s g_{n,s,t} - d_{n,t} = \sum_l K_{n,l} \cdot f_{l,t} \quad \forall \quad n,t,\\ 
{g}^-_{n,s,t} \cdot G_{n,s} \leq g_{n,s,t} \leq {g}^+_{n,s,t} \cdot G_{n,s} \quad \forall \quad n,t.\\ 
\end{split}
\end{align}
In the equation for the nodal power balance, $d_{n,t}$ is the demand at node $n$ and time $t$,  $K_{n,l}$ the network's incidence matrix and $f_{l,t}$ the actual power flow on line $l$ at time $t$. The energy generation in the dispatch constraint is limited downwards by its minimal generation potential ${g}^-_{n,s,t}$ and limited upwards by its maximal generation potential ${g}^+_{n,s,t}$, where $G_{n,s}$ describes the capacity of a carrier type $s$ at node $n$. 

Storage units obey additional constraints given by the state of charge (soc) equations, which describe the charging and discharging behaviour:
\begin{align}
\begin{split}
\mathrm{soc}_{n,s,t} &= \eta_0 \cdot \mathrm{soc}_{n,s,t-1} + \eta_1 \cdot g_{n,s,t,\textrm{store}} \\&- \eta_2^{-1} \cdot g_{n,s,t,\textrm{dispatch}}\\ &+ \mathrm{inflow}_{n,s,t} - \mathrm{spillage}_{n,s,t} \quad \forall \quad n, s, t > 1.\\
\end{split}
\end{align}
Charging efficiencies are described by $\eta$, where $\eta_0$ describes standing losses (neglected in the following), $\eta_1$ efficiencies of storage uptake and $\eta_2$ efficiencies of storage dispatch (both assumed as 0.9 for battery storage). The term $\mathrm{inflow}$ covers any external inflow, e.g. energy inflow into hydro reservoirs and $\mathrm{spillage}$ losses for instance by reaching maximum storage capacity. 

Active power flows $f_l$ on a transmission link $l$ are constrained by the maximum transmission capacity $F_l$ of the considered link:
\begin{align}
\begin{split}
|f_l\left(t\right)| &\leq F_l \quad \forall \quad l\\
\end{split}
\end{align}
In this study, every link-capacity was set to 5 GW.

We use the software toolbox Python for Power System Analysis (PyPSA, \cite{brown2017pypsa}) to perform the optimisation.
This toolbox has been widely used in the research community, for instance to study the interplay of sector coupling and transmission grid extensions \cite{brown2018synergies} or the impact of climate change on a future European power system \cite{schlott2018impact}.

\subsection{Data and Assumptions}
Renewable generation potentials for the weather year 2013 were obtained from renewables.ninja \cite{pfenninger2016renewables}. Wind potentials are based on wind speeds from the MERRA-2 reanalysis \cite{gelaro2017modern} and the current wind turbine fleet per country. Solar potentials are based on CM-SAF's SARAH satellite dataset \cite{muller2015surface} with default configurations.

Hydro inflow and capacity data were obtained from the RESTORE 2050 project \cite{alexander_kies_2017_804244}.
Hydro energy was assumed to be fully exploited in Europe and hence not extendable. Hydro energy inflow was modelled using a potential energy approach and surface runoff data \cite{kies2016restore}.

Consumption data per country were obtained from the open power system data project \cite{opsd}.

The cost assumptions for all considered technologies were derived from different sources (\cite{carlsson2014etri,schroeder2013current}), they were annualised with a discount rate of 7\% and are given in Table \ref{tab:costsassumptions}.

\begin{table*}[!t]
\begin{center}
\resizebox{.6\textwidth}{!}{\begin{tabular}{ lrrrrrr }

\hline
Technology & Capital Cost & Marginal Cost & Emission \\ 
           &  [mu/GW/a] & [mu/MWh] & CO$_2$ [ton/MWh] \\ 
\hline
OCGT & 49,400 & 58.385 & 0.635\\
Wind & 127,450 & 0.010 & 0\\
PV & 61,550 & 0.010 & 0 \\
Coal & 145,000 & 25.000 & 1.0 \\
Hydro Reservoir & 0 & 0 & 0\\
Battery & 120,389  & 0 & 0\\
\hline
\end{tabular}}
\caption{Annualised cost assumptions for generation and storage technologies derived from different sources \cite{carlsson2014etri,schroeder2013current}. All energy quantities refer to exergy values.}
\label{tab:costsassumptions}
\end{center}
\end{table*} 

\section{Results}
We cost-optimise the problem stated in Eq. (3) - (6)
for a carbon price between 0 and 250 mu/ton and a distribution parameter $\alpha$ between 0 and 2.

\begin{figure}[!h]
    \centering
    \includegraphics[width=.49\textwidth]{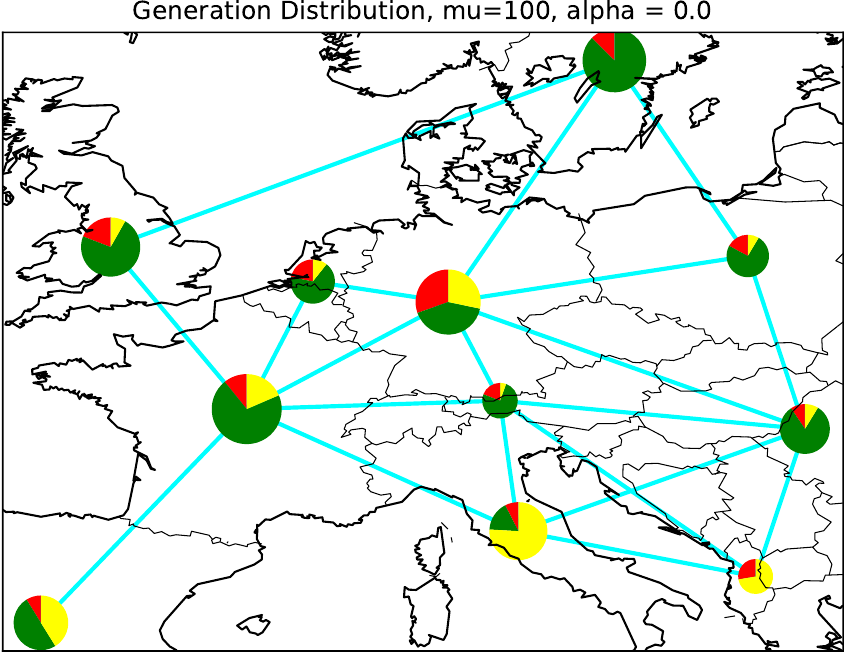}
    \caption{Distribution of energy generation from the different sources wind (green), solar PV (yellow), OCGT (red) and coal (blue) for $\bar{\mu}=100$ and $\alpha = 0.0$. Sizes of the circles indicate overall energy generation at the single nodes.}
    \label{fig:distribution0}
\end{figure}

\begin{figure}[!h]
    \centering
    \includegraphics[width=.49\textwidth]{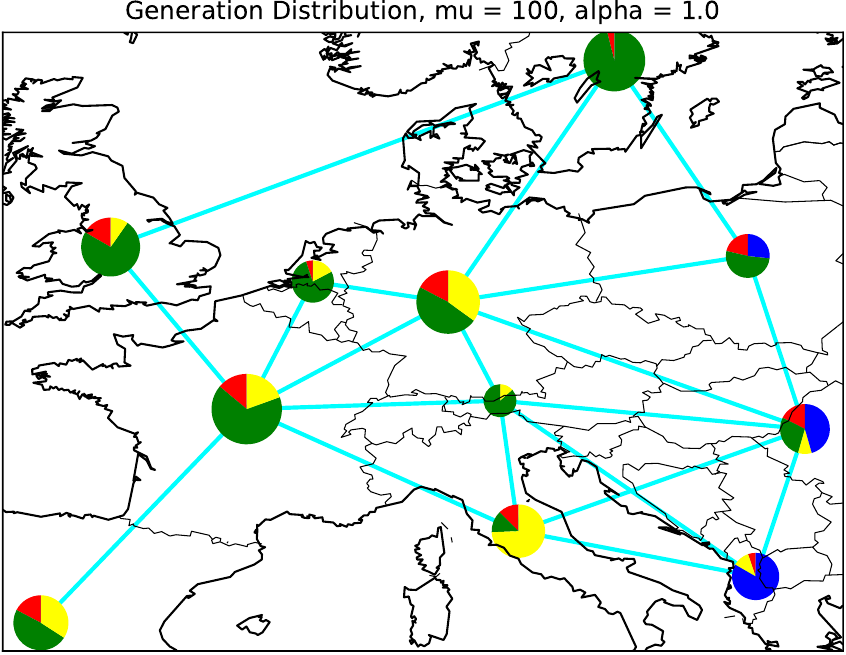}
    \caption{Distribution of energy generation from the different sources wind (green), solar PV (yellow), OCGT (red) and coal (blue) for $\bar{\mu}=100$ and $\alpha = 1.0$. Sizes of the circles indicate overall energy generation at the single nodes.}
    \label{fig:distribution1}
\end{figure}

\begin{figure}[!h]
    \centering
    \includegraphics[width=.49\textwidth]{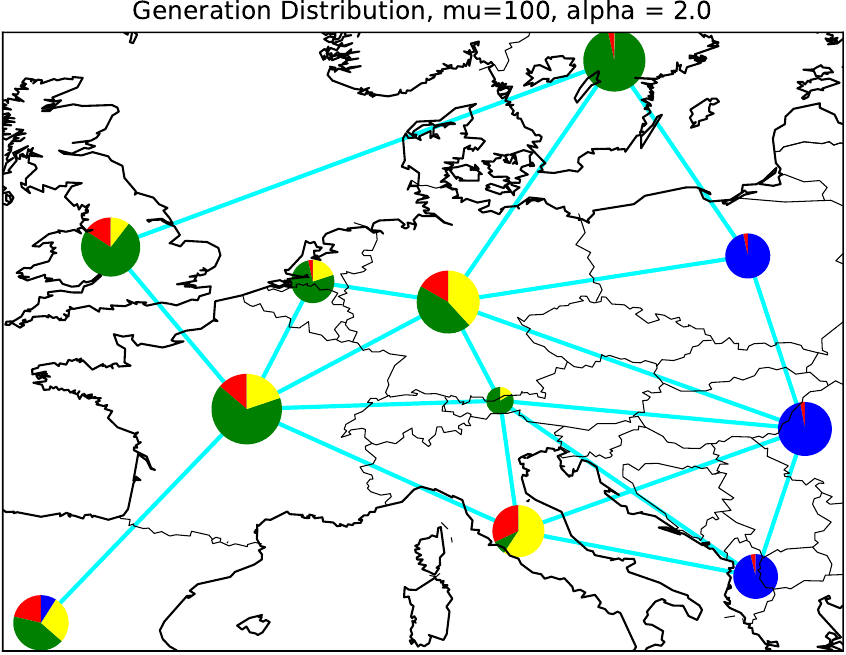}
    \caption{Distribution of energy generation from the different sources wind (green), solar PV (yellow), OCGT (red) and coal (blue) for $\bar{\mu}=100$ and $\alpha = 2.0$. Sizes of the circles indicate overall energy generation at the single nodes.}
    \label{fig:distribution2}
\end{figure}

Fig. \ref{fig:distribution0} - \ref{fig:distribution2} show the European distribution of dispatched energy for a carbon price of 100 mu/ton and
distribution parameters $\alpha$ equal to zero (Fig. \ref{fig:distribution0}), one (Fig. \ref{fig:distribution1}), and two (Fig. \ref{fig:distribution2}). At a first glance, several apparent aspects remain the same: most nodes are producing from the very same generation sources.
However, for an uneven carbon price distribution, i.e. increasing the distribution parameter $\alpha$, a large share of coal generation remains in Eastern Europe. At $\alpha=2.0$, OCGT has also replaced a relevant share of solar PV in the southern node 6.
In the south-eastern node 11, coal replaces solar PV almost entirely, whereas in node 10 coal replaces a large proportion of wind 
and generates approximately 50\% of energy. For $\alpha=1$, local carbon prices remain low at 32 mu/ton (node 11) and 38 mu/ton (node 10), thus keeping coal a competitive source of energy generation. For $\alpha=2.0$, where the south-eastern nodes experience no carbon prices, these effects are even stronger emphasised.

\begin{figure}[!h]
    \centering
    \includegraphics[width=.5\textwidth]{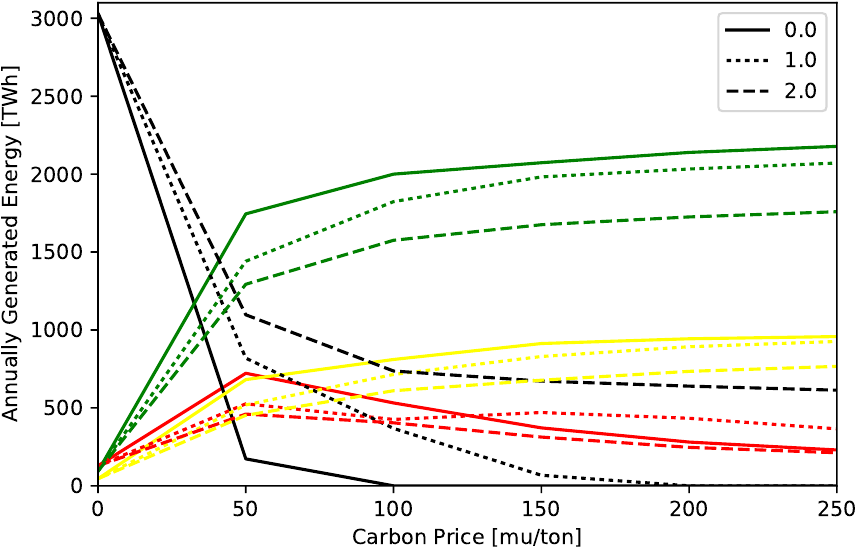}
    \caption{Annually generated energy of renewable (green: wind, yellow: solar PV) and conventional (red: OCGT, black: coal) generation technologies in dependency of the carbon price and the distribution parameter $\alpha$ shown in the legend.}
    \label{fig:dispatch}
\end{figure}

\begin{figure}[!h]
    \centering
    \includegraphics[width=.5\textwidth]{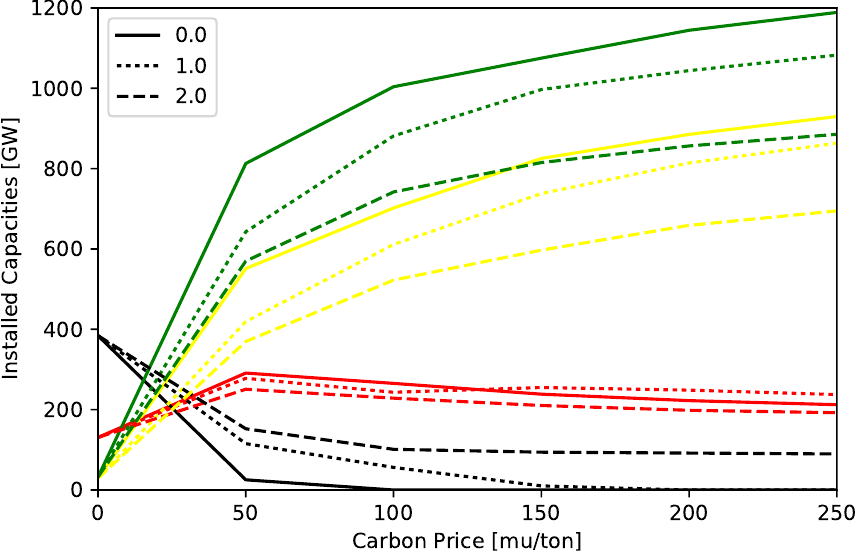}
    \caption{Capacities of renewable (green: wind, yellow: solar PV) and conventional (red: OCGT, black: coal) generation technologies in dependency of the carbon price and the distribution parameter $\alpha$ shown in the legend.}
    \label{fig:capacities}
\end{figure}

The general evolution of dispatch and capacities per generation technology are shown in Fig. \ref{fig:dispatch} and \ref{fig:capacities}.
Without carbon prices, coal is a competitive source of generation that dominates. However, even at a rather limited base price of 50 mu/ton, the coal share has vanished almost entirely in the homogeneous scenario. If carbon prices are distributed unevenly, a share of coal survives up until higher base prices or even entirely up to 250 mu/ton. Another interesting observation is the dispatch of OCGT being higher at a base price of 50 mu/ton for $\alpha=0.0$ than for $\alpha=1$ or $2$, but its share is overtaken soon thereafter.

\begin{figure}[!h]
    \centering
    \includegraphics[width=.5\textwidth]{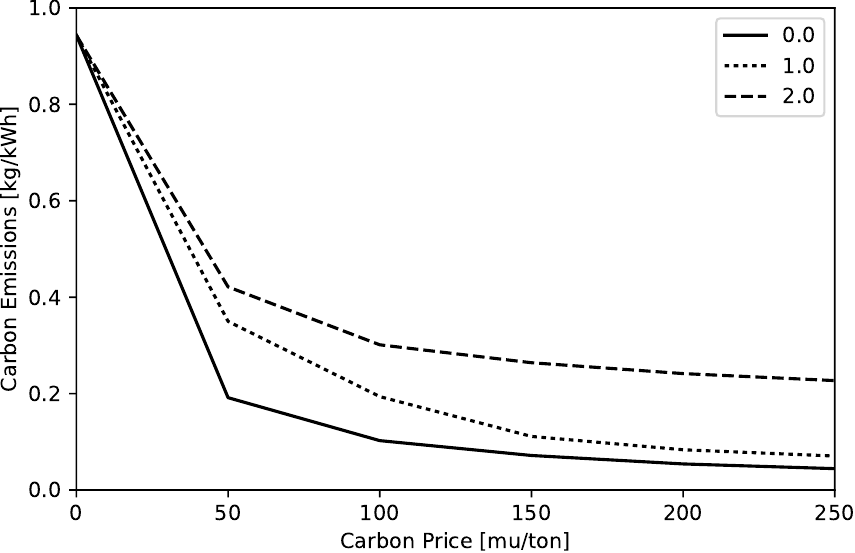}
    \caption{Total CO$_2$ emissions for variations of carbon price and distribution parameter.}
    \label{fig:emissions}
\end{figure}

Carbon leakage can be observed in the curves of CO$_2$ emissions in Fig. \ref{fig:emissions}.
At a base CO$_2$ price of 50 mu/ton,
a strongly inhomogeneous CO$_2$ price distribution leads to an increase of carbon dioxide emissions of 100\% compared to the homogeneous case.
At a CO$_2$ price of 100 mu/ton and beyond, this difference is even stronger emphasised.
For a base price of 250 mu/ton, CO$_2$ prices vary between 79 and 413 mu/ton at $\alpha=1$, while at $\alpha=2$ they vary between 0 and 577 mu/ton.
Although transmission capacities are limited, vast carbon leakage occurs and prevents a deep decarbonisation. 

\begin{figure}[!h]
    \centering
    \includegraphics[width=.5\textwidth]{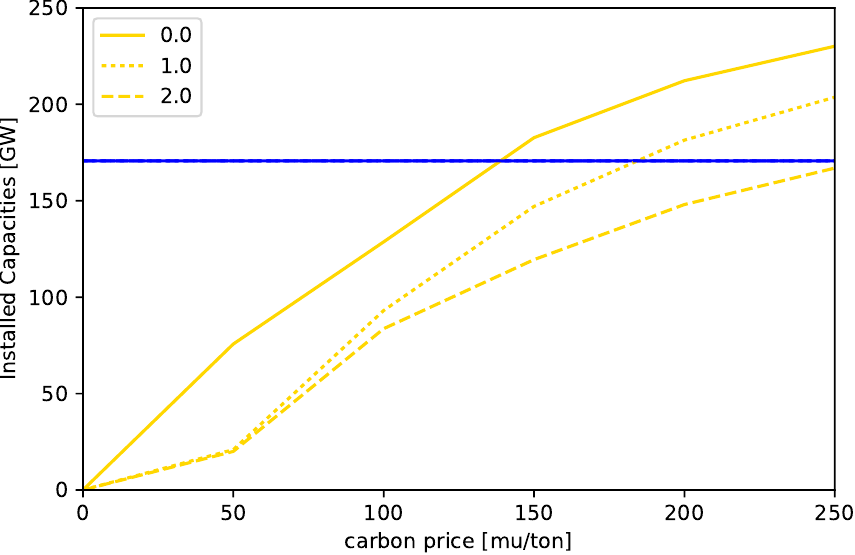}
    \caption{Installed hydro (blue) and battery (yellow) storage capacities in dependency
of the carbon price and the distribution parameter shown in the legend.}
    \label{fig:storage}
\end{figure}

The increasing shares of renewables for homogeneous carbon price distributions and growing carbon base prices go hand in hand with increasing installations of battery storage (Fig. \ref{fig:storage}). While hydro power is kept constant throughout the model runs, battery capacities are added and surpass the installed capacities of hydro power for high carbon prices. For battery storage, a ratio of installed energy to power capacity of 8 hours has been assumed. Installed capacities of 125 GW are thus equivalent to an installed energy capacity of 1 TWh.
Compared to hydro power, which is mostly located in the regions 1 and 7 and has a overall energy capacity of more than 130 TWh, this is neglible. However, battery storage can be installed at every region and can also be used to take energy up from renewable sources, whereas hydro power only has a natural inflow from surface runoff in the model.

\begin{figure}[!h]
    \centering
    \includegraphics[width=.5\textwidth]{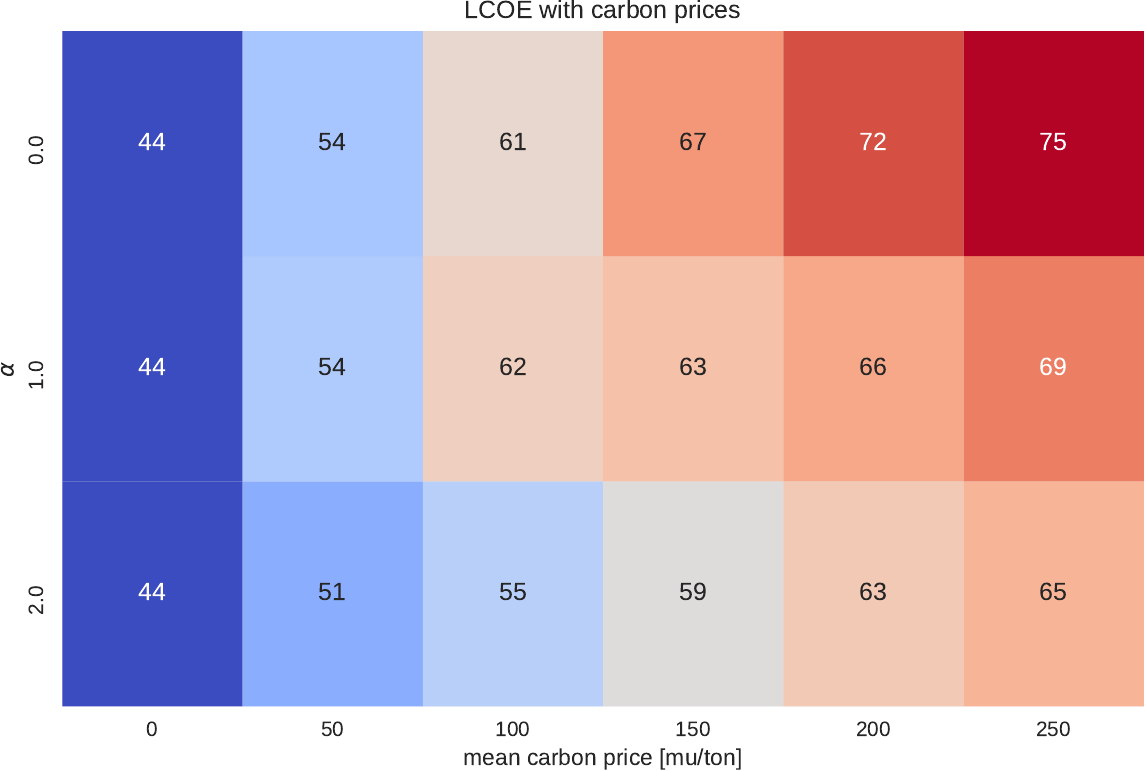}
    \caption{Levelised cost of electricity in mu/MWh including carbon prices as part of marginal generation costs.}
    \label{fig:lcoe}
\end{figure}

\begin{figure}[!h]
    \centering
    \includegraphics[width=.5\textwidth]{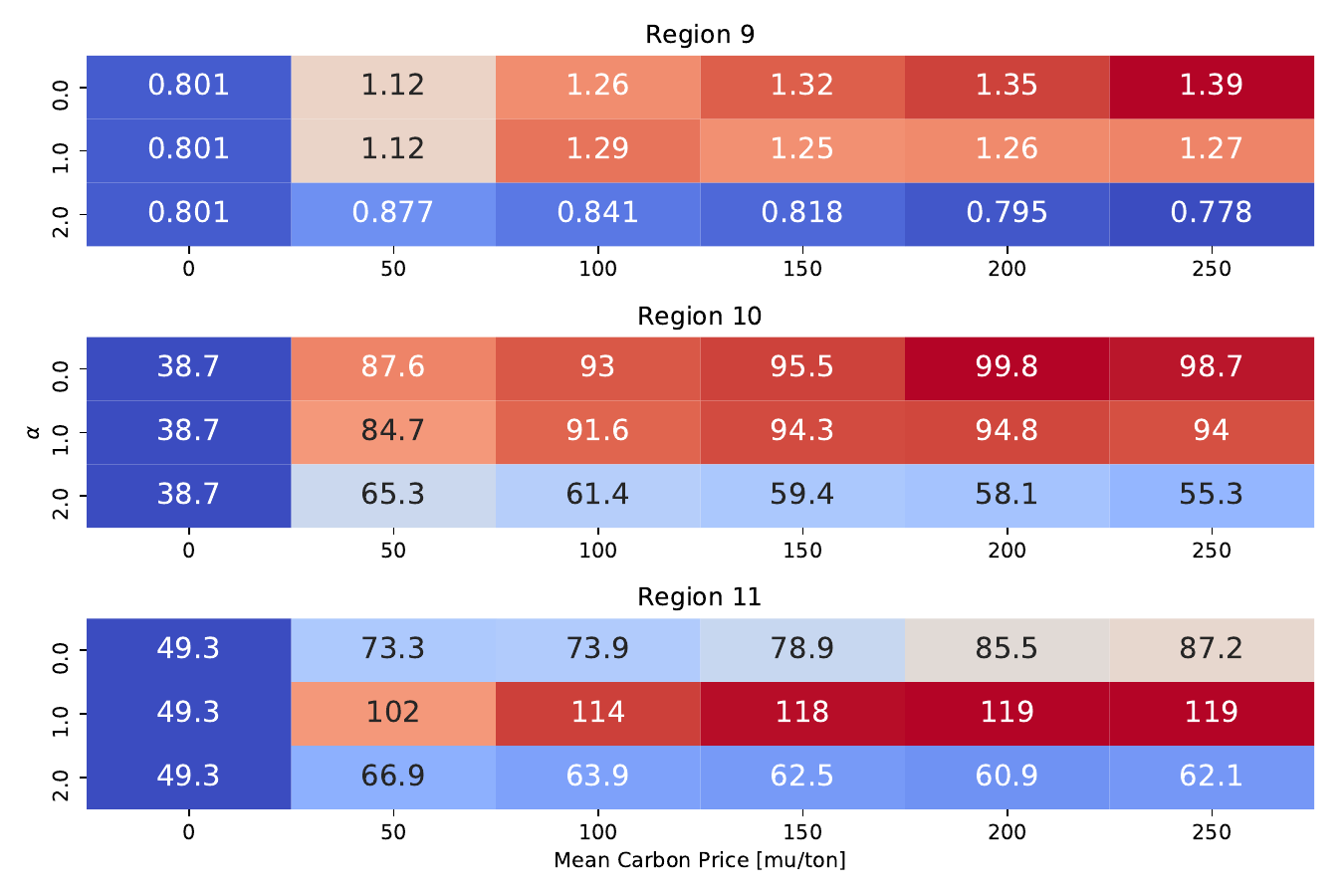}
    \caption{System costs for exemplary regions divided by the demand of the respective region [mu/MWh].}
    \label{fig:sc}
\end{figure}

System cost including carbon prices are lowest for every base scenario in the scenario of inhomogeneous distribution among countries (Fig. \ref{fig:lcoe}). However, for low base prices of 50 or 100 mu/ton, the homogeneous distribution remains cheaper than the medium scenario ($\alpha=1$). This is potentially due to OCGT being more competitive as a generation technology in the homogeneous scenarios in the high-GDP western and northern countries, thus allowing for relevant cost savings.
Additional transmission expansion potentially could wear off this effect. This effect can also be observed in Fig. \ref{fig:sc}. It shows the costs of regions 9-11 in Eastern Europe including costs of generation, dispatch, storage and carbon prices divided by the demand of the specific region. For region 11, the medium scenarios are by far those, where most costs accumulate. The underlying processes can be concluded from Fig. \ref{fig:distribution0} - \ref{fig:distribution2}:
For $\alpha = 0$, this region supplies its own demand with some limited exchange with the neighbours. For $\alpha = 1$, overall dispatch from generators in this region grows significantly leading to the rise in cost. For $\alpha = 2$, overall dispatch shrinks again slightly and solar PV vanishes despite good conditions in south-eastern Europe. However, since the share of solar PV grows in this case also in region 6, it is likely that limited transmission capacities and the strongly correlated solar feed-in in different neighbouring regions make the cost-efficient use of solar PV in region 11 impossible and drive down overall cost at this node.


\section{Summary, Conclusions and Outlook}
In this paper, it has been investigated how inhomogeneous carbon prices affect a European power system. It has been shown that varying CO$_2$ prices can lead to significant carbon leakage.
Assuming CO$_2$ prices to increase monotonously with increasing GDP per capita reflecting economic strength of a region, has led to waste amounts of coal generation remaining in Eastern and South-Eastern Europe and consequently carbon leakage from the West to the East. 
The effect of the distribution parameter on overall system cost has been less clear, indicating the involvement of several opposing processes.

It can be concluded that a significant difference in carbon prices across European countries can lead to significant carbon leakage and therefore endanger greenhouse gas emission reduction targets.

Several extensions of this work are straightforward:
\begin{itemize}
    \item Schyska et al. \cite{schyska2019implications} have recently shown that varying capital cost for investments in renewables among European countries (they strongly vary because of economic and political differences) have profound impact on the distribution of renewables and should be considered in studies on the European power system.
It is an reasonable assumption that the effect of inhomogeneous capital cost and inhomogeneous carbon prices oppose each other. This effect could be combined with varying CO$_2$ prices in future work to study their complex interaction.
\item The electricity sector is only a small part of the entire energy system and consumes only around a quarter of primary energy demand in Europe.
Another question worth investigation is in such a setting as the one presented, what carbon leakage occurs in a sector-coupled energy model, where sectors might or might not be fully electrified. As Brown et al. \cite{brown2018synergies} have shown, such a system is also easier to decarbonise, if flexibility potentials of the heating and transport sector can be utilized. 
\item As an alternative to an all-electric world as it has been considered in this paper, power-to-x technologies are often debated as an alternative. The major energy carriers in power-to-x scenarios are hydrogen and methane produced via electrolysis and Sabatier processes. These offer additional flexibilites, especially if the heating and transport sector are investigated, as well, and can be taken into account in future work.
\item In the model presented in this paper, the energy system is build up from scratch (greenfield investment problem). However, taking the existing power plant park into account (brownfield investment problem) and performing a pathway optimization could lead to different results. In such a setting, decomissioning dates of power plants would be considered as well as limited investment budgets per time period (e.g. per annum). This would also allow to investigate policy control mechanisms and potentially model the interactions between policies, the energy system and decarbonisation targets.
\end{itemize}

\section*{Acknowledgment}
This work is financially supported by the CoNDyNet II project (Bundesministerium fuer Bildung und Forschung, Fkz. 03EK3055C).
and the Net-Allok project (Bundesministerium fuer Wissenschaft und Energie, Fkz. 03ET4046A).
The authors thank Martin Greiner (Aarhus) for helpful suggestions and comments to improve this paper.

\bibliographystyle{IEEEtran}
\bibliography{sample}
\end{document}